\documentclass[twocolumn,showpacs,preprintnumbers]{elsart5p}
\journal{Physics Letters A}
\usepackage{ifpdf}
\usepackage{graphicx,amssymb,lineno,amsmath}
\ifpdf
\usepackage[%
  pdftitle={Instructions for use of the document class
    elsart},%
  pdfauthor={Simon Pepping},%
  pdfsubject={The preprint document class elsart},%
  pdfkeywords={instructions for use, elsart, document class},%
  pdfstartview=FitH,%
  bookmarks=true,%
  bookmarksopen=true,%
  breaklinks=true,%
  colorlinks=true,%
  linkcolor=blue,anchorcolor=blue,%
  citecolor=blue,filecolor=blue,%
  menucolor=blue,pagecolor=blue,%
  urlcolor=blue]{hyperref}
\else
\usepackage[%
  breaklinks=true,%
  colorlinks=true,%
  linkcolor=blue,anchorcolor=blue,%
  citecolor=blue,filecolor=blue,%
  menucolor=blue,pagecolor=blue,%
  urlcolor=blue]{hyperref}
\fi

\makeatletter
\def\elsartstyle{%
    \def\normalsize{\@setfontsize\normalsize\@xiipt{14.5}}
    \def\small{\@setfontsize\small\@xipt{13.6}}
    \let\footnotesize=\small
    \def\large{\@setfontsize\large\@xivpt{18}}
    \def\Large{\@setfontsize\Large\@xviipt{22}}
    \skip\@mpfootins = 18\p@ \@plus 2\p@
    \normalsize
} \@ifundefined{square}{}{} \makeatother

\newcommand{\ket}[1]{|#1\rangle}
\newcommand{\mi}{\mathrm{i}}

\begin{document}

\begin{frontmatter}
\title{Detuned Electromagnetically Induced Transparency in $N$-type Atom System}

\author[GUO]{Bin Luo},
\author[GUO]{Xiao Li}
\and\author[GUO]{Hong Guo\corauthref{cor}}
\corauth[cor]{Corresponding author. Phone: +86-10-6275-7035, Fax:
+86-10-6275-3208.} \ead{hongguo@pku.edu.cn}
\address[GUO]{CREAM Group, School of
Electronics Engineering and Computer Science and School of Earth and
Space Science,\\
 Peking University, Beijing 100871, P. R. China}

\begin{abstract}
The electromagnetically induced transparency (EIT) in an $N$
configuration is studied under both resonant and off-resonant
conditions. In a certain off-resonant condition the dark state of
the four level system, which is almost the same as the resonant dark
state in $\Lambda$ configuration, is rebuilt. The actual system with
damping is examined using optical Bloch equation, both numerically
and analytically. Based on this detuned dark state, some new
applications with frequency shifts can be realized.
\end{abstract}
\end{frontmatter}
Recently, manipulations of the refraction and absorption properties
by means of quantum coherence and interference (QCI) has been
studied extensively for resonant interaction between laser and
three-level atom ensembles with $\Lambda$-, $V$- and $\Xi$-type
configurations  and show the effects of electromagnetically induced
transparency (EIT) \cite{harris1997eit}, electromagnetically induced
absorption (EIA) \cite{akulshin1998eia,lezama1999eia}, and etc.
These effects lead to possibilities of the manipulation of light
pulse group velocity, and the subluminal and superluminal
propagation of light pulse have been demonstrated
experimentally\,\cite{fulton1995cwe}.

It is straightforward to consider the possibility of adjusting an
external driving field into this three-level system to modulate the
absorption and dispersion properties. The interaction between light
pulse and four level system is also studied for $N$
configurations\,\cite{taichenachev1999eia}, with the observations of
the nonlinearity at low light level\,\cite{hoonsoo2003olk}, as well
as the interchange between subluminal and superluminal
propagation\,\cite{han2005sas}.

Other important applications of EIT, especially in the $\Lambda$
configuration, are light storage and quantum memory. The possibility
of light storage based on EIT is proposed first by Fleischhauer and
Lukin\,\cite{PhysRevA.65.022314,fleischhauer2000dsp} with the dark
state polariton model and experimentally realized in ultracold
sodium atoms\,\cite{liu2001oco}, in hot rubidium
vapor\,\cite{phillips2001sla}, and in
solids\,\cite{PhysRevLett.88.023602}. For quantum memory, single
photon storage has been realized both in room-temperature atomic
gas\,\cite{signelhot} and in cold atom cloud\,\cite{signelcold}.

As is well known that, in a strict sense, the $\Xi$- and $V$-type
three-level schemes do not show EIT owing to the absence of a
meta-stable dark state\,\cite{fleischhauer}. Dark state is very
important in applications of EIT, especially in light storage.
However, in those studies about $N$ configuration, no exact dark
state is achieved in resonant condition. At the same time, in a
$\Lambda$ system where a dark state condition is satisfied, the
external control is impossible and its single transparency window
brings some limitations in applications, especially in real
communications.

\begin{figure}
\centering
\includegraphics[width=6cm]{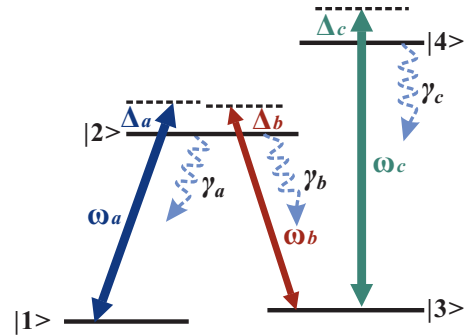}
\caption{(color online)\,The configuration of $N$-type four level
atom. $|1\rangle$ and $|3\rangle$ are two ground levels, $|2\rangle$
and $|4\rangle$ are excited states, and the corresponding damping
rates are $\gamma_{a,b,c}$. Levels $|1\rangle$ and $|2\rangle$, and
$|3\rangle$ and $|4\rangle$ are coupled by strong light, i.e.,
coupling light and driving light, respectively, while $|2\rangle$
and $|3\rangle$ are coupled by a weak probing
light.}\label{case1of4}\end{figure}

Here, an $N$ configuration atomic system is studied in both resonant
and off-resonant conditions. In some off-resonant conditions, dark
states can be reachieved, with more than one transparency windows.
Hence, some new applications under those off-resonant conditions are
proposed.

\section{Detuned Dark States}

Let us consider an $N$-type atom system shown in Fig.
\ref{case1of4}, the Hamiltonian of which (under the rotating wave
frame) can be written as

\noindent
\begin{equation}\label{Hamiltonian 1}
\begin{aligned}
\hat{H}&=\hbar\Delta_a|1\rangle\langle1|+\hbar\Delta_b|3\rangle\langle3|+\hbar(\Delta_b-\Delta_c)|4\rangle\langle4|\\
&-\frac{\hbar}{2}(\Omega^*_a|1\rangle\langle2|+\Omega^*_b|3\rangle\langle2|+\Omega^*_c|3\rangle\langle4|+\mathrm{H.c.}),
\end{aligned}
\end{equation}

where
$\Omega_{a}=-2\mathbf{d}_{21}\cdot\hat{\epsilon}_a\mathcal{E}_a/\hbar$,
$\Omega_{b}=-2\mathbf{d}_{23}\cdot\hat{\epsilon}_b\mathcal{E}_b/\hbar$,
and
$\Omega_{c}=-2\mathbf{d}_{43}\cdot\hat{\epsilon}_c\mathcal{E}_c/\hbar$
are Rabi frequencies. $\mathbf{d}$ is the dipole moment,
$\hat{\epsilon}_i\;(i=a,b,c)$ is the unit vector, $\mathcal{E}_i$ is
the envelop of the incident light and $\omega_i\;(i=a,b,c)$ is the
central frequency.

A general expression of the eigenstates of the Hamiltonian is very
complicated. For simplicity, the coupling and driving field are set
to be resonant, while the situation of $\Delta_b=0,\pm \Omega_c/2$
will be examined.

If the driving light $\Omega_c$ is not introduced, Fig.
\ref{case1of4} shows exactly a $\Lambda$ configuration with dark
state

\noindent
\begin{equation}\label{dark state_lambda}
\ket{\Psi^0}=\cos\theta\ket{1}-\sin\ket{3},
\end{equation}

\noindent where
$$\theta=\tan^{-1}\left(\frac{\Omega_a}{\Omega_b}\right),$$ with
eigenvalue 0\,\cite{RevModPhys.70.1003}.

However, when applying the driving field $\Omega_c$, the dark state
in resonant condition vanishes. Eigenstates of the Hamiltonian then
become

\noindent
\begin{equation}\centering
|\Psi_i\rangle=\frac{1}{\sqrt{2}}(\cos\theta_1|1\rangle-\sin\theta_1|3\rangle+\cos\theta_2|2\rangle-\sin\theta_2|4\rangle),\label{resonant
condition}
\end{equation}

where
$$\theta_1=\tan^{-1}\left(\frac{\Omega_a\Omega_b}{\Omega_b^2+\Omega_c^2-4\lambda_i^2}\right)
,\;\;\theta_2=\tan^{-1}\left(\frac{\Omega_c\Omega_b}{\Omega_c^2-4\lambda_i^2}\right).$$
and $\hbar\lambda_i,(i=1,2,3,4)$ are the eigenvalues of the
Hamiltonian, satisfying the equation:

\noindent
\begin{equation}\label{eigenrs}
\left( 2\lambda  - {{\Omega }_a} \right)
  \left( 2\lambda  + {{\Omega }_a} \right)
  \left( 4{\lambda }^2 - {{{\Omega }_b}}^2 - {{{\Omega }_c}}^2
    \right)=\Omega_a^2\Omega_b^2.\nonumber
\end{equation}

Compared with Eq. (\ref{dark state_lambda}), state shown in Eq.
(\ref{resonant condition}) is the superposition of all four levels
and hence is not a dark state at all. The driving field introduces
state shifts and destroys the resonant dark state. However, Eq.
(\ref{resonant condition}) shows some new properties.

It is apparent that there is no strict dark state in this resonant
condition, although in Eq. (\ref{resonant condition}) some amplitude
of the states can be very small. $\theta_1$ and $\theta_2$ show that
the population are transferring between two lower levels and two
upper levels independently, each shares a half. So, more than EIT,
some applications related to population transfer is available here.

However, when $\Delta_b=\pm\Omega_c/2$, one of the eigenstates of
the Hamiltonian becomes

\noindent
\begin{equation}\label{dark1}
|\Psi_c^{\pm}\rangle=-\frac{\Omega_b}{\Omega_a}|1\rangle+|3\rangle\pm|4\rangle,
\end{equation}

\noindent with eigenenergy 0, while the other three eigenstates are

\noindent
\begin{equation}
\begin{aligned}
|\Psi^{\pm}_i\rangle &=-2\Omega_a(\lambda_i\mp\Omega_c)|1\rangle+4\lambda_i(\lambda\mp\Omega_c)|2\rangle\nonumber\\
&-\Omega_b(2\lambda_i\mp\Omega_c)|3\rangle+\Omega_b\Omega_c|4\rangle.\nonumber
\end{aligned}
\end{equation}

\noindent with $\lambda^{\pm}_i,(i=1,2,3)$ satisfying
$$8\lambda^3\mp8\Omega_c\lambda^2-(2\Omega_a^2+2\Omega_b^2)\lambda\pm(2\Omega_a^2+\Omega_b^2)\Omega_c=0.$$

The state shown in Eq. (\ref{dark1}) is the same as in $\Lambda$
configuration except $|4\rangle$ term. It has no contribution from
$|2\rangle$ and therefore is a dark state, which implies that if the
atom is prepared in this state, the probability of the transition to
$|2\rangle$ is zero and subsequently, there is no absorption. Thus,
the medium is transparent to the probing light with detuning
$\Delta_b=\pm\Omega_c/2$.
\section{Modulations by the driving light}
The driving field is introduced in order to modulate the
$\Lambda$-type EIT properties. With both the coupling and driving
fields detunings being zero, the detuned dark states
[Eq.\,(\ref{dark1})] have been derived. Next, we will study the
influence on this modulation of both the detuning and intensity,
which is represented by Rabi frequency of the driving field here.

The coupling light is assumed to be resonant for the maximum
$\Lambda$-type EIT signal. In this case, it is evident that the
eigenstate with zero eigenvalue of the Hamiltonian
[Eq.\,(\ref{Hamiltonian 1})] is dark for the probing light, since
there is no energy exchange between the probing light and the atom.
The Hamiltonian reaches its zero eigenvalue when

\noindent
\begin{equation}\label{eigenvalue zero}
    \Delta_b^{\pm}=\frac{1}{2}(\Delta_c\pm\sqrt{\Delta_c^2+\Omega_c^2}).
\end{equation}

And the case $\Delta_c=0$, $\Delta_b=\pm\Omega_c/2$ is exactly the
same as derived in previous section, and correspondingly, the
detuned dark states are

\noindent
\begin{equation}\label{ds 2}
    |D^{\pm}\rangle=\Omega_b|1\rangle-\Omega_a|3\rangle+\frac{\displaystyle \Omega_c}{\displaystyle
    \Delta_c\mp\sqrt{\Delta_c^2+\Omega_c^2}}|4\rangle.
\end{equation}

According to this result, the influence of the driving field can be
understood as following, while $\Delta_c$ is assumed to be positive
for convenience: (i) If the driving field is resonant, the detuned
effects are the same on both sides of the detunings so that a
symmetrical detuned EIT can be drawn. (ii) When the driving field is
not resonant, or if it is far detuned with $\Delta_c\gg\Omega_c$,
then the dark state can be achieved at $\Delta_b\simeq0$ and
$\Delta_b\simeq\Delta_c$. However, it is no longer symmetric because
when $\Delta_b\simeq0$, the state remains
$\Omega_b|1\rangle-\Omega_a|3\rangle$, which is exactly the dark
state in three-level $\Lambda$-type EIT profile; but nearly all the
population will be pumped to the level $|4\rangle$ if
$\Delta_b\simeq\Delta_c$ and merely a EIT signal will be available.

\section{The dressed state picture}
The detuned dark state [Eq.\,(\ref{ds 2})] can be understood more
clearly in a dressed state picture. Since the influence of the
driving field is interesting here, the states dressed only by the
driving field is sufficiently evident for analysis.
\begin{figure}[h]
\centering
\includegraphics[width=3.6 in]{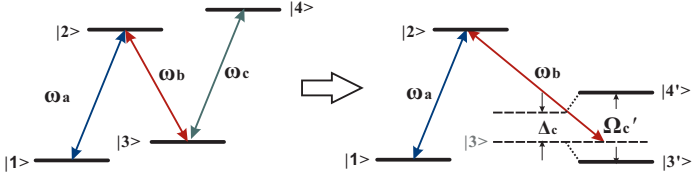}
\caption{(color online)\,The dressed state picture with states
dressed by the driving field.
$\Omega_c'=\sqrt{\Delta_c^2+\Omega_c^2}$. Apparently it can be
understood as two $\Lambda$-type EIT system with level $|3\rangle$
shifted. }\label{dressed}
\end{figure}


According to the dressed state theory, the driving field, with no
influence on $|1\rangle$ and $|2\rangle$, dressed the states
$|3\rangle$ and $|4\rangle$ into $|3'\rangle$ and $|4'\rangle$,
respectively,

\noindent
\begin{equation}
\begin{aligned}
  |3'\rangle &=& \cos \vartheta |3\rangle-\sin \vartheta |4\rangle,\nonumber \\
  |4'\rangle &=& \sin \vartheta |3\rangle+\cos \vartheta |4\rangle,\nonumber
  \end{aligned}
\end{equation}

\noindent where $\tan 2\vartheta=-\Omega_c/\Delta_c$, and the energy
shifted is exactly the same as that in Eq. (\ref{eigenvalue zero}),
as illustrated in Fig. \ref{dressed}. Thus, the detuned EIT effect
can, in this case, be viewed as an Autler-Townes splitting in an EIT
medium.

From the above analysis, the detuning of the driving field is set to
be zero for symmetric detuned dark states.
\section{The density matrix approach with spontaneous dampings included and double transparency windows}
In the above analysis, no dampings are included. However, in
realistic situations, dampings should be considered, especially for
the case of the detuned dark state [Eqs.\,(\ref{dark1}) or (\ref{ds
2})] derived previously, in which an excited state $|4\rangle$ is
contained. In this case, the spontaneous emission of $|4\rangle$
will surely affect the transparency feastures.

\begin{figure}[!htb]
\centering
\includegraphics[width=3.6 in]{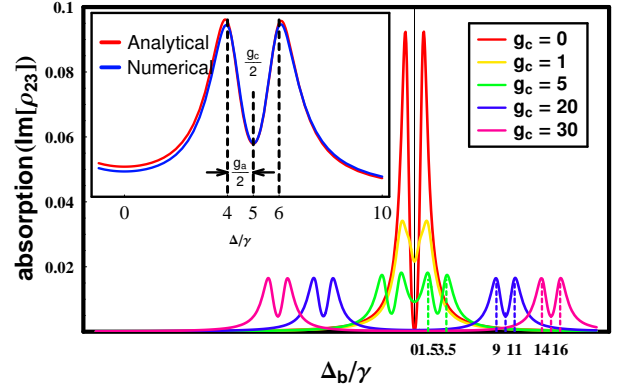}
\caption{(color online)\,Absorption modulated by
$g_c=\Omega_c/\gamma$. $g_a=\Omega_a/\gamma=2$,
$g_b=\Omega_b/\gamma=0.2$, $\Delta_a=\Delta_c=0$. When $g_c$ is
arbitrarily large ($\gg 5$), the absorption line is detuned by $g_c$
without changing its shape. All the damping rates are assumed to be
equal here for simplicity ($\gamma_{a,b,c}=\gamma$). The inset is
the comparison of numerical and analytical solutions at $g_b=0.2$,
$g_a=2$ and $g_c=10$. }\label{case1of4vary1}
\end{figure}

For simplicity, we assume that there exists only radiative damping,
for the collision-induced dampings can be well controlled in cold
atoms, so the coherence dephasing rates are $\Gamma_{12} =
\frac{1}{2}(\gamma_a+\gamma_b)$,
   $\Gamma_{13} =  0$,
   $\Gamma_{14} = \frac{1}{2}\gamma_c$,
  $ \Gamma_{23} = \frac{1}{2}(\gamma_a+\gamma_b)$,
  $ \Gamma_{24} = \frac{1}{2}(\gamma_a+\gamma_b+\gamma_c)$, and
  $ \Gamma_{34} = \frac{1}{2}\gamma_c$.
To see the absorption and dispersion characteristics of the probing
light in such a system, we give numerically the steady state
solution of $\rho_{23}$, as illustrated in Fig. \ref{case1of4vary1}.
It is shown that the pumping light generates one (when $\Omega_c =
0$) or more (when $\Omega_c \neq 0$) transparency windows, which is
very evident when $\Omega_c$ is large.

Next, an analytical solution is given. For simplicity, all the
damping rates are assumed to be equal ($\gamma_{a,b,c}=\gamma$), and
$\Delta_a$ and $\Delta_c$ are set to be zero for resonant pumping
lights. If the requirement for laser energy in $\Lambda$ system is
satisfied\,\cite{PhysRevLett.74.2447}, i.e. $\Omega_a\gg\Omega_b$,
one has

\noindent
\begin{equation}\label{Neit} \begin{aligned}
\rho_{23}=&\frac{g_b}{2g_c(1+2g_c^2)}\big([A(\delta_+)+B(\delta_+)]E(\delta_-)\notag\\
&+[A(\delta_-)-B(\delta_-)]E(\delta_+)-[E(\delta_+)+E(\delta_-)]\big)\notag\\
\big/&\big(\,E(\delta_+)E(\delta_-)+ \frac{16}{
g_c^2}[E(\delta_+)-E(\delta_-)]^2\,\big),\end{aligned}
\end{equation}

\noindent where

\noindent
    \begin{equation}
    \begin{aligned}
    A(x) &=-g_c(1+g_c^2)(4x+\mi),\\
    B(x) &=(1-2\mi x)(1-g_c^2),\\
    E(x) &=1-2g_a^2+2(\mi+2x)(2\mi+2x).
    \end{aligned}
    \end{equation}

 \noindent
  Here, $\delta_\pm=\delta_b\mp\frac{\displaystyle g_c}{\displaystyle 2}$
denotes the positive and negative frequency shifts, relative to
two-photon resonance. $g_i=\Omega_i/\gamma,\;(i=a,b,c)$ and
$\delta_b=\Delta_b/\gamma$ represent the normalized Rabi frequencies
and probing light detunings, respectively.

The interest here is focused on the order of the magnitude of $g_c$,
which is proportional to the square root of the intensity of the
coupling light. If $g_c\gg 1$, $B$ can be much smaller than $A$ and
$\frac{\displaystyle 16}{\displaystyle
g_c^2}[E(\delta_+)-E(\delta_-)]^2$ can be ignored. Then the result
becomes

\noindent
\begin{equation}\label{sim2}
\rho_{23}=\frac{\displaystyle 1}{\displaystyle
2}[f(\delta_+)+f(\delta_-)],
\end{equation}

\noindent where $f(x)=-\frac{\displaystyle g_b}{\displaystyle
2}\frac{\displaystyle \mi+4\delta_b}{\displaystyle E(x)}$, as
illustrated in the inset of Fig. \ref{case1of4vary1}. The numerical
results are also shown in the inset of Fig. \ref{case1of4vary1} for
comparison.

From Fig. \ref{case1of4vary1}, one finds that there are three
transparent windows at $\pm\Omega_c/2$ detunings and resonance. As
mentioned above, the dark states generate the detuned transparency
and the resonant transparency can be achieved directly from Eq.
(\ref{resonant condition}). Since $\Omega_b$ is relatively small, we
have $4\lambda^2\approx\omega_c^2$ and $\theta_2\approx\pi/2$, which
makes $|2\rangle$ in Eq. (\ref{resonant condition}) vanish and leads
to a dark-like state and therefore the transparency.

The numerical solution shows that if the shift is large enough, say,
about $10\gamma$, all the transparency windows are separated
evidently. Experimental conditions often guarantee the assumption we
have made to be valid.

The EIT element $f$ is important  since, from which, one finds that
the three lights play different roles: $g_c$ determines the
transparency shift, $g_a$ determines the line shape and width, and
$g_b$ is a separate term which only affects the amplitude. This
property makes it easy for manipulation of the state and the
coherence.

For further consideration, the dark state [shown in Eq.
(\ref{dark1})] is of great interest since it totally agrees with the
dark state in $\Lambda$ configuration, which is of great importance
in many respects. It should be emphasized that the driving Rabi
frequency $\Omega_c$ has the impact only on the frequency shift of
probing light but has nothing to do with the dark state itself.
Also, since the amplitude is the same between $|3\rangle$ and
$|4\rangle$, there is always enough population, more than half, to
interact with the light. So, the detuned EIT looks like a perfect
shift of the $\Lambda$-type resonant EIT scheme since even the dark
states have been shifted to both sides, and the techniques, such as
the adiabatic evolution, the dark state polariton and those based on
the dark state, can be used here in a symmetric frequency condition.
\section{Further Analysis for $\gamma_c$}
However, from the above analysis, one may find out that the
absorption at $\pm \Omega_c/2$ is not zero. This is because only
$|2\rangle$ is dark, while $|4\rangle$, which keeps damping out, is
still populated. If the dampings are included, the states expressed
in Eq. (\ref{dark1}) are not the ideal dark-states and it has an
absorption determined by $\gamma_c$.
\begin{figure}[!htb] \centering
\includegraphics[width=8.6cm]{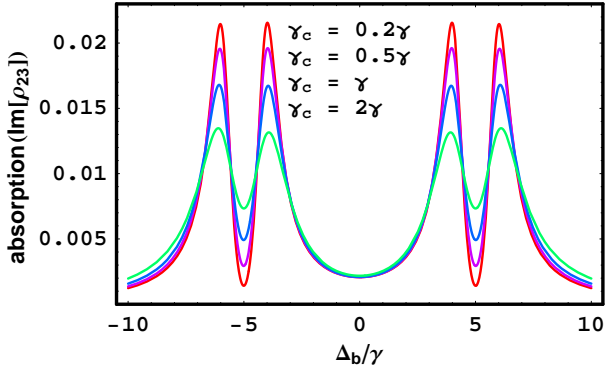}
\caption{(color online)\,The influence of different $\gamma_c$.
Here, $g_a=2\gamma$, $g_b=0.2\gamma$, $g_c=10\gamma$. Both the
coupling and driving lights are resonant. $\gamma_a=\gamma_b=\gamma$
and $\gamma_c/\gamma=0.2,0.5,1,2$.}\label{double_lambda}
\end{figure}

\begin{figure}[h] \centering
\includegraphics[width=8.6cm]{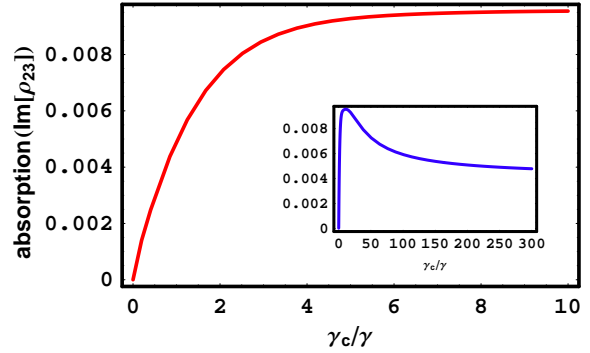}
\caption{(color online)\,The relation between absorption at $\pm
\Omega_c/2$ and $\gamma_c$. Here, $g_a=2\gamma$, $g_b=0.2\gamma$,
$g_c=10\gamma$, $\gamma_a=\gamma_b=\gamma$. $\delta_b=g_c=5$. Both
the coupling and driving lights are
resonant.}\label{absoptionshifted}
\end{figure}

The influence of $\gamma_c$ is illustrated in Fig.
\ref{double_lambda} and \ref{absoptionshifted}, which show that the
absorption is proportional to $\gamma_c$ when $\gamma_c$ is small.

From the above analysis, it is clear that the damping rate from
$|4\rangle$,\,i.e., $\gamma_c$, plays an important role. If
$\gamma_c$ is approximately equal to $\gamma$, the expressions have
been given already. If $\gamma_c$ is small compared to $\gamma$, as
required experimentally, the imaginary part of $\rho_{23}$ can be
derived as

\noindent
\begin{equation}\label{absym01}
\mathrm{Im}[\rho_{23}(\delta_b=0)]=\frac{\zeta \,{\kappa }^2\,{g_a}}
  {4\,{\kappa }^2 + {\left(  {\kappa }^2 -1\right)
  }^2\,{{g_a}}^2},\nonumber
\end{equation}

\noindent
 where $\kappa={g_c}/{g_a},\;\zeta={g_b}/{g_a}.$ This
approximate expression is derived under the condition $\kappa\geq
1,\;g_a\geq1, \zeta\ll1$ and $\gamma_c\leq \zeta\gamma $, which
requires a strong coupling light and an even stronger driving light,
compared with a weak probing light and a small damping rate. If all
the conditions are satisfied, the absorption [$\delta_b=\pm(g_c\pm
g_a)/2$] and transparency ($\delta_b=\pm g_c/2$) can be derived in
the same way, which are

\noindent
$$\mathrm{Im}[\rho_{23}]^{\textrm{trans}}=\frac{\varGamma \,\zeta \,\left[ 12\,{\kappa }^2 +
      \left( 1 - 3\,{\kappa }^2 + 4\,{\kappa }^4 \right) \,{{g_a}}^2
      \right]
    }{32\,{\kappa }^2\,{g_a} +
    2\,{\left( 1 - 4\,{\kappa }^2 \right) }^2\,{{g_a}}^3},$$
$$\mathrm{Im}[\rho_{23}]^{\textrm{absorp}}=\frac{\varGamma \,\zeta \,\left[ 2\,{\left( 1 + 2\,\kappa  \right) }^2 +
      4\,{\left( \kappa  + {\kappa }^2 \right) }^2\,{{g_a}}^2 \right] }{8\,
    {\left( \kappa  + {\kappa }^2 \right) }^2\,{g_a}\,
    \left( 4\,\varGamma  + {\zeta }^2\,{{g_a}}^2 \right) },$$

\noindent where $\varGamma=\gamma_c/\gamma$. It is apparent that
when the probing light is resonant,  the absorption is independent
of $\gamma_c$. But when the probing light is detuned to be around
$\pm
\Omega_c/2$, the absorption characteristics strongly depends on $\gamma_c$. 

\begin{figure}[!htb]
 \centering
\includegraphics[scale=0.35]{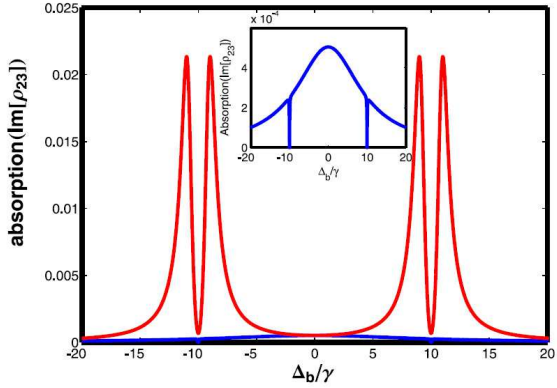}
\caption{(color online)\,Illustration of the collapse of
transparency windows. Here, $\Omega _a/\gamma= 2$,\,$\Omega
_b/\gamma=\text{0.2}$,\, $\Omega_c/\gamma= 20$. Both the coupling
and driving lights are resonant. (red: $\gamma_c$ = 0.1 case; blue:
$\gamma_c$ = 10$^{-7}$ case; inset: enlarged figure of $\gamma_c$ =
10$^{-7}$ case.) We find that when $\gamma_{c}$ is reduced to
$10^{-7}$, the transparency window has collapsed.
}\label{specialplot.eps}
\end{figure}

However, it should be noted that the level $\ket{4}$ should not be
too stable, as illustrated in Fig.\;\ref{specialplot.eps}. In the
limit $\gamma_{c}\rightarrow0$, the relationship between the
absorption and $\Delta_b$ would undergo great changes (Fig.
\ref{specialplot.eps}): The absorption at all frequencies has been
reduced to less than two percent of that at $\gamma_{c} = 0.1$. That
is to say, when one probing light injects into this system, the
system is almost transparent, so the difference between transparency
window and other frequencies can be neglected. In this sense, when
level $\ket{4}$ is too stable, the windows we create would collapse.
Moreover, the overall property of absorption curve (inset of Fig.
\ref{specialplot.eps}) is also significantly changed.

\section{Discussion and Conclusion}

In conclusion, we apply a driving optical field to the $\Lambda$
configuration and destroy the dark state in resonant condition.
However, at $\pm\Omega_c/2$ detunings of the probing field, the dark
states are rebuilt with almost complete fidelity. The influences of
the three lights and destructive effects by dampings are discussed.

For experimental realization, a closed four-level N-type atom system
is required. One of this kind of atom configurations is available in
the D2 line of $^{133}$Cs. The four levels can be chosen, according
to respective dipole moments, as: $|6^2S_{1/2},
F=3\rangle\rightarrow|1\rangle$, $|6^2P_{3/2},
F=4\rangle\rightarrow|2\rangle$, $|6^2S_{1/2},
F=4\rangle\rightarrow|3\rangle$ and $|6^2P_{3/2},
F=5\rangle\rightarrow|4\rangle$. In this configuration, the four
levels compose a closed system and all the damping rates are equal
($\gamma_a=\gamma_b=\gamma_c=2\pi\times5.22\mathrm{MHz}$). Of
course, $\gamma_c$ is not chosen to be sufficiently small here to
restrain the absorption, as required by the above theoretical
analysis. However, compared with the experimental results in V-type
and cascade configuration EIT experiments\,\cite{fulton1995cwe},
this problem is not crucial for a light pulse storage experiment.
But one has to notice that for single photon storage, it might bring
much more disadvantages.

This phenomenon can, hopefully, be widely used in many areas of
quantum optics and quantum information, such as slow light, light
storage and quantum memory, based on the current well-founded
quantum manipulation techniques. In concrete, it may give a
possibility of storing two light pulse with two  different
frequencies using techniques based on dark state polariton
\cite{fleischhauer2000dsp}, which gives potential possibility for
wavelength division multiplexing \cite{Agrawal}.

This work is supported by the National Natural Science Foundation of
China (Grant No. 10474004), National Key Basic Research Program
(Grant No. 2006CB921401) and DAAD exchange program: D/05/06972
Projektbezogener Personenaustausch mit China (Germany/China Joint
Research Program).


\clearpage
\end{document}